\documentclass{article}
\usepackage[english]{babel}
\usepackage{spconf,amsmath,epsfig}
\usepackage{amsthm}
\usepackage{amsfonts}
\usepackage{amssymb}
\usepackage{aas_macros}
\usepackage[ruled]{algorithm2e}

\newcommand{\lnd}{\mathrm{LN}}
\newcommand{\eqi}{\gamma}

\newcommand{\Tr}{^\mathrm{\small T}}

\newcommand{\va}{\alpha}

\newcommand{\Sm}{\mathrm{S}}
\newcommand{\Wm}{\mathrm{W}}

\newcommand{\Id}{\boldsymbol{\mathrm{I}}}

\newcommand{\Kmb}{\boldsymbol{\mathrm{K}}}

\newcommand{\Mmb}{\boldsymbol{\mathrm{M}}}

\newcommand{\Phb}{\boldsymbol{\Phi}}

\newcommand{\norm}[1]{\left\lVert #1 \right\rVert}

\newcommand{\abs}[1]{\left\lvert #1 \right\rvert}

\newcommand{\T}{\mathrm{F}}

\newcommand{\Cc}{\mathcal{C}}

\newcommand{\Nc}{\mathcal{N}}

\newcommand{\Pc}{\mathcal{P}}

\newcommand{\Prj}{\mathcal{P}}

\newcommand{\Rn}{\mathbb{R}^n}

\DeclareMathOperator{\prox}{prox}

\renewcommand{\ge}{\geqslant}
\renewcommand{\le}{\leqslant}

\theoremstyle{plain}
\newtheorem{theorem}{\textbf{Theorem}} %[section]

\newtheorem{lemma}[theorem]{\textbf{Lemma}}

\newtheorem{propo}[theorem]{\textbf{Proposition}}

\newtheorem{definition}[theorem]{\textbf{Definition}}

\title{Data augmentation for galaxy density map reconstruction\footnote{Submitted to ICIP 2011 on the 01/21/11}}
\name{F.-X. Dup\'e$^\text{a}$, M.J. Fadili$^\text{b}$ and J.-L. Starck$^\text{a}$}
\address{\small \begin{tabular}{cc}
    \begin{minipage}{0.45\linewidth}\begin{center} $^\text{a}$ AIM UMR CNRS - CEA \\ 91191 Gif-sur-Yvette France \end{center}\end{minipage} & 
    \begin{minipage}{0.45\linewidth}\begin{center} $^\text{b}$ GREYC UMR CNRS 6072 \\ Universit\'e de Caen Basse-Normandie \\ ENSICAEN \\ 14050 Caen France\end{center} \end{minipage} \\
  \end{tabular}
  \vspace*{-0.5cm}
}
\begin{document}
%\ninept
%
\maketitle
\begin{abstract}
  The matter density is an important knowledge for today cosmology as many phenomena are linked to matter
  fluctuations. However, this density is not directly available, but estimated through lensing maps or galaxy surveys.
  In this article, we focus on galaxy surveys which are incomplete and noisy observations of the galaxy
  density. Incomplete, as part of the sky is unobserved or unreliable. Noisy as they are count maps degraded by Poisson
  noise. Using a data augmentation method, we propose a two-step method for recovering the density map, one step for
  inferring missing data and one for estimating of the density. The results show that the missing areas are efficiently
  inferred and the statistical properties of the maps are very well preserved.
\end{abstract}
\section{Introduction}
\label{sec:intro}

Information about the origin of the Universe is encoded inside the cosmological matter distribution. It is an
important challenge to be able to estimate such a distribution. However, the whole matter is not available, only a
biased observation is possible using count maps estimator of massive objects. These maps are degraded both by shot noise
and astrophysics phenomena (e.g. Milky Way, galactic dust).

Current methods for reconstructing the density maps are more focused on the denoising problem than on the missing data.
For example, \cite{Kitaura2009:wiener} propose a Wiener filter for estimating the SDSS DR6 survey
\cite{Adelman-McCarthy2008:sdss}. A maximum a posteriori was introduced in \cite{Kitaura2010:poisson} with a Poisson
data fidelity term (including the mask operator) and a log-normal prior.

This last prior comes from Hubble who found in 1934 that the distribution of galaxy counts is well fitted by a
log-normal distribution. It was later confirmed by others studies that this statement is correct for a given interval of
scales (mostly the medium scales)
\cite{Coles:1991,Kitaura2010:poisson}. % On large scale, the matter distribution is supposed to be mostly
% Gaussian, while on small scales, the distribution begins to diverge largely from the log-normal distribution to a
% complex one.

In this paper, we propose both to estimate the density map and infer the missing data using a data augmentation process
\cite{Tanner1987:data}. First, we present how to generate realistic data with a random texture synthesis
algorithm. Secondly, a maximum a posteriori estimator is proposed for the density using both the log-normal prior and a
sparse prior. Then, we apply our method on real data and compare the results with how state-of-art alternatives in the
literature.

% \subsection*{Notation}
% \label{sec:notation}

% We also denote by $\opnorm{\Mmb}= \max_{\vx \neq 0} \frac{\norm{\Mmb\vx}}{\norm{\vx}}$
% the spectral norm of $\Mmb$.  $\vx$ and $\va$ are respectively reordered vectors of
% image samples and transform coefficients.

\subsection*{Notation and terminology} %{Sparse image representation}
\label{sec:sparse-image-repr}

We denote by $\norm{.}_2$ the norm associated with the inner product in $\Rn$, and $\Id$ is the identity operator on $\Rn$. A function $f$ is coercive, if $\lim_{\norm{\delta}_2 \rightarrow +\infty}f\left(\delta\right)=+\infty$.  $\Gamma_0(\Rn)$ is the class of all proper lower semi-continuous convex functions from $\Rn$ to $]-\infty,+\infty]$. 

Let $\delta \in \Rn$ be an $n$-pixel map. $\delta$ can be written as the superposition of elementary atoms
$\varphi_\gamma$ parametrized by $\gamma \in \mathcal{I}$ such that $\delta = \sum_{\gamma \in \mathcal{I}}
\alpha_\gamma \varphi_\gamma = \Phb \va,\quad \abs{\mathcal{I}} = L, ~ L\ge n$.  We denote by $\Phb$ the dictionary
i.e. the $n\times L$ matrix whose columns are the generating waveforms $\left(\varphi_\gamma\right)_{\gamma \in
  \mathcal{I}}$ all normalized to a unit $\ell_2$-norm.  The forward transform is the non-necessarily
square matrix $\T = \Phb\Tr \in \mathbb{R}^{L\times n}$.  A dictionary matrix $\Phb$ is said to be a frame with bounds
$c_1$ and $c_2$, $0<c_1\le c_2<+\infty$, if $ c_1 \norm{\delta}^2 \le \norm{\Phb\Tr \delta}^2 \le c_2 \norm{\delta}^2$~.  A
frame is tight when $c_1=c_2=c$, i.e. $\Phb\Phb\Tr =c\Id$.  In the rest of the paper, $\Phb$ will be an orthobasis or a
tight frame with constant $c$.

\section{Recovering the galaxy density}
\label{sec:recov-galaxy-dens}

The galaxy surveys are count maps degraded by Poisson noise and parts of the sky are not subject to interferences with
astrophysical phenomena and objects (e.g. Milky Way, galactic dust). The underlying image formation model is,
\begin{equation}
  \label{eq:1}
  y \sim \Pc(\Mmb\bar{m}(1+\delta))~,
\end{equation}
where $y$ is the observation, $\delta$ the density field, $\bar{m}$ the mean number of counts (e.g. galaxies) per pixel
and $\Mmb$ the binary mask operator (i.e. 0 where data are missing and 1 else).

Estimating $\delta$ from $y$ is an ill-posed problem, so priors are needed to reduce the solution set.  The density
field is assumed to follow a log-normal distribution \cite{Coles:1991}, i.e. $(1+\delta) \sim \mathcal{LN}(\mu,\Sigma) \sim
\exp(\Nc(\mu,\Sigma))$, where the mean $\mu$ and the covariance matrix $\Sigma$ are the parameters of the underlying
Gaussian field.

% For the theory, we know that the mean of $\delta$ should be null and its power spectra $\Sm$ can be computed from a
% given model of the Universe (through cosmological parameters). As the spherical harmonic domain is a Karhunen–Lo\`eve
% basis for our signal, the covariance matrix (of the underlying normal field) is linked to the power spectra, $\Sigma =
% \log(1 + \Sm)$ and the mean $\mu = -\diag(\Sm)^2$. Using this knowledge, we propose to build a texture generating
% process for inferring the missing data.

\subsection{A data augmentation method}
\label{sec:inpa-using-data}

Inferring missing data is a longstanding and delicate problem. Most used methods in statistics to handle missing data are the expectation-maximization (EM) algorithm and multiple imputation (MI); see \cite{Little1987} for a comprehensive review.

As the distribution of $\delta$ is known, the data augmentation method~\cite{Tanner1987:data} seems to be the most adapted. This is an EM scheme (described in Algorithm~\ref{algo:dataaug}) where, first, the missing data are inferred using multiple imputations (i.e. several \textit{new} observations are generated at each iteration). Then, the M-step
consists in estimating the sought after parameter(s) based on the complete formed data.

\begin{algorithm}[h]
  \small
  \noindent{\bf{Task:}} Data augmentation method for both inferring the missing data and estimating the parameters. \\
  \noindent{\bf{Parameters:}} The observation $y$, number of iterations
  $N_{\mathrm{iter}}$ and the number of imputations $N_{\mathrm{MI}}$. \\
 \noindent{\bf{Main iteration:}} \\
 \noindent{\bf{Initialization:}} Compute estimates of the log-normal parameters, $\hat{\Sigma}$ and $\hat{\mu}$, and
 of the mean number of counts $\hat{\bar{m}}$.
  \noindent{\bf{For}} $t=0$ {\bf{to}} $N_{\mathrm{iter}}-1$,
  \begin{itemize}
    \setlength{\topsep}{0pt}
    \setlength{\parskip}{0pt}
    \setlength{\itemsep}{0pt}
    \setlength{\partopsep}{0pt}
  \item \underline{E-step}: create $N_{\mathrm{MI}}$ complete observations by filling the missing area using the prior
    distribution of the density field and the estimated parameters,
  \item \underline{M-step}: for each complete observation estimate the density field, the log-normal parameters,
    $\Sigma_i$ and $\mu_i$ of the field and the mean $\bar{m}_i$,
  \item \underline{Update step}: each parameter is updated using the estimates from the multiple imputations, $\hat{\mu}
    = (\sum_i \mu_i)/N_{\mathrm{MI}}$, $\hat{\Sigma} = (\sum_i \Sigma_i)/N_{\mathrm{MI}}$, $\hat{\bar{m}} = (\sum_i
    \bar{m}_i)/N_{\mathrm{MI}}$.
  \end{itemize}
  \noindent{\bf{End main iteration}} \\
 \caption{The data augmentation scheme.}
  \label{algo:dataaug}
\end{algorithm}

The multiple imputation is very useful when parameters are updated, and the number of imputations is linked to the amount
of missing data (for example 5 imputations should be sufficient for $0.5$ ratio~\cite{Little1987}). The tricky point
remains the generation of realistic data.  With prior information, one can advocate Markov Chain Monte Carlo (MCMC)
methods, but such methods are usually computationally very expensive.

\subsection{Generating realistic data (E-step)}
\label{sec:gener-real-data}

As an alternative to MCMC methods, we propose a texture synthesis-like method for creating realistic data inside the missing areas obeying the appropriate statistical properties underlying the image formation model. Indeed, such data should respect the model formation~\eqref{eq:1} (with $\Mmb \equiv \Id$) and the density field has to follow a given log-normal distribution.

Inspired by the work of \cite{Portilla2000} in texture synthesis, our statistical data generation can be cast as a hard feasibility problem,
\begin{equation}
  \label{eq:2}
  \text{find}\quad {\delta \in \cap_{i=1}^3\ \Cc_i}~, 
\end{equation}
where each $\Cc_i$ represents a constraint set. In our case, we want to constraint the mean and the covariance of the
underlying Gaussian field (as log-normality is assumed), but the observed parts of the density field must be
preserved. The solution is then computed by projecting the data onto the constraints sets using the Von-Neumann
alternating projections algorithm,
\begin{equation}
  \label{eq:7}
  \delta_{t+1} =  \Prj_{\Cc_1} \circ \Prj_{\Cc_2} \circ \Prj_{\Cc_3} ( \delta_t)~,
\end{equation}
where $\Cc_1$ is the convex constraint set pertaining to the observed part preservation, $\Cc_2$ is associated to the
covariance constraint (non-convex), and $\Cc_3$ is the mean constraint set (convex). Algorithm~\ref{algo:texture}
summarizes the steps of this the synthesis method for our problem. As $\Cc_2$ is not convex, sophisticated arguments are
needed to potentially prove convergence of the sequence $(\delta_t)_{t\in\mathbb{N}}$ to a point in $\cap_{i=1}^3\
\Cc_i$ (if non-empty). This will be left to a future work. In practice, only a few iterations were necessary to produce
satisfactory results.

\begin{algorithm}[h]
  \small
  \noindent{\bf{Task:}} Generating realistic data to infer the missing data. \\
  \noindent{\bf{Parameters:}} The current estimate of the density field $\hat{\delta}$, the observation $y$, the binary
  mask operator $\Mmb$, the covariance matrix $\Sigma$, the mean $\mu$, the mean number of count $\bar{m}$,
  and the number of iterations $N_{\mathrm{tex}}$. \\
  \noindent{\bf{Initialization:}} $p_0 = \hat{\delta}$, \\
  Replace the data in the missing areas of $p$ with random data using the log-normal distribution.\\
 \noindent{\bf{Main iteration:}} \\
 \noindent{\bf{For}} $t=0$ {\bf{to}} $N_{\mathrm{tex}}-1$,
  \begin{itemize}
    \setlength{\topsep}{0pt}
    \setlength{\parskip}{0pt}
    \setlength{\itemsep}{0pt}
    \setlength{\partopsep}{0pt}
  \item \underline{Get the Gaussian field}: $z_t = \log(1+p_t)$,
  \item \underline{Estimate the mean}: $m^z_t = \mathbb{E}(z_t)$,
  \item \underline{Contraint the mean}: $z_t = z_t - m^z_t + \mu$, 
  \item \underline{Estimate the covariance}: $\Sm^z_t = \mathrm{Cov}(z_t)$,
  \item \underline{Constraint the covariance}: $z_t = \Sigma^{\tfrac{1}{2}} \Sm_z^{-\tfrac{1}{2}} z_t$,
  \item Update the estimate: $p_t = \Mmb \hat{\delta} + (\Id - \Mmb) (\exp(z_t)-1)$.
  \end{itemize}
  \noindent{\bf{End main iteration}} \\
  \underline{Add Poisson noise}: $\hat{p} \sim \Pc(\bar{m}(1+p_{N_{\mathrm{tex}}}))$.\\
 \noindent{\bf{Output:}} The imputed observation $\hat{y} = \Mmb y + (\Id-\Mmb)\hat{p}$.
  \caption{Texture generating process for the imputation step (E-step).}
  \label{algo:texture}
\end{algorithm}

\subsection{Estimating the galaxy density (M-step)}
\label{sec:estim-galaxy-dens}

Now, we assume complete observations,
\begin{equation}
  \label{eq:11}
  \hat{y} \sim \mathcal{P}\left(\bar{m}(1+\delta)\right)~.
\end{equation}
The density field estimation amounts now to a Poisson denoising problem. By adopting a Bayesian framework and using a
standard maximum a posteriori (MAP) rule, we combine data fidelity with both log-normal prior and sparsity prior.

% By adopting a Bayesian framework and using a
% standard maximum a posteriori (MAP) rule, our goal is to minimize the following functional with respect to the
% representation coefficients $\va$: {\small
%   \begin{gather}
%     \label{eq:6}
%     (\mathrm{P}_{\lambda,\gamma,\psi}): \min_{\va\in\mathbb{R}^L} J(\alpha) \\
%     J : \alpha \mapsto  -\ell\ell(y|.)\circ\Phb(\va) + \gamma q_{LN}(.|\mu,\Sigma) \circ \Phb(\va)+ \lambda \sum_{i=1}^{L} \psi(\va[i])~, \nonumber
%   \end{gather}
% } with $\delta = \Phb\va$, $-\ell\ell$ the data fidelity term, $q_{\lnd}$ the log-normal prior and its weighting
% parameter $\gamma$, $\Psi = \sum_i \psi$ the sparsity-penalty and $\lambda$ the regularization parameter. Notice that,
% we implicitly assume that the $\va[i], 0\le i \le n$ are independent and identically distributed.

% Using the Bayesian paradigm, we use the maximum a posteriori approach for estimating the density field.  Then, using a
% synthesis prior, the objective function (denoted $J$) is composed of three terms: the data fidelity term (from the
% Poisson likelihood) with two regularization terms for log-normal prior and the sparsity constraints,

The data fidelity term is directly constructed from the anti log-likelihood of the multivariate Poisson distribution,
{\footnotesize
\begin{align}
  \label{eq:9}
  -\ell\ell\ &: \eta \in \mathbb{R}^n \mapsto \sum_{i=1}^n f_{\mathrm{poisson}}(\eta[i]), \\
  \text{if } y[i] > 0,\quad
  f_{\mathrm{poisson}}(\eta[i]) &=
  \begin{cases}
    -y[i] \log(\eta[i]) + \eta[i] & \text{if } \eta[i] > 0,\\
    +\infty & \text{otherwise,}
  \end{cases} \nonumber \\
  \text{if } y[i] = 0,\quad
  f_{\mathrm{poisson}}(\eta[i]) &=
  \begin{cases}
    \eta[i] & \text{if } \eta[i] \in [0,+\infty), \\
    +\infty & \text{otherwise.}
  \end{cases} \nonumber
\end{align}
}
where $\eta = 1 + \delta$.

% {\small
%   \begin{align}
%     \label{eq:8}
%     -\ell\ell(y|\delta) &= \sum_{i=0}^{n-1} \bar{m}(1 + \delta[i]) - y[i]\log(\bar{m}(1 + \delta[i])) \\
%     \notag &+ \log(y[i]!),\quad  \forall i,\ \delta[i] > -1~.
%   \end{align}
% }
% Notice that this term is convex, but its gradient is not Lipschitz (i.e. its norm is not bound) and so can not be
% directly use inside a gradient descent scheme. However, there exist technics for dealing properly with such
% difficulty~\textbf{citer Figueiredo, Moi et Chaux ?}. We will later propose another way to avoid such contraint.

The regularization term for the log-normal prior is given by anti-log likelihood of the multivariate log-normal
distribution for a covariance matrix $\Sigma$ and a mean $\mu$, {\small
  \begin{gather}
    \label{eq:4}
   q_{\lnd}(\delta|\mu,\Sigma) = \\  \frac{1}{2} (\log(1+\delta) - \mu)\Tr \Sigma^{-1} (\log(1+\delta) - \mu) 
    + \sum_{i=1}^n \log(1 + \delta[i])~. \nonumber
  \end{gather}
} Notice that $q_{\lnd}$ is not convex because the $\log$ function is concave.

% Remark, the $\ell_2$ norm over the logarithm of the density field may lead to an over-smoothing of the structures if the
% equilibrium parameter $\gamma$ is too high.  Secondly,
% Moreover, the optimal value of such parameter remains an open question (see \cite{Giryes2010} for a example of finding
% such parameter).

\subsubsection{The optimization problem}

The non-convexity of $q_{\lnd}$ can be avoided using a change of variable, $z = \log(1 + \delta)$ and assuming that the
underlying Gaussian field is sparse inside the dictionary domain. Then, the new optimization problem, with $z=\Phb\va$,
is, {\footnotesize
  \begin{gather}
    \label{eq:3}
    (\mathrm{P}_{\lambda,\gamma,\psi}): \min_{\alpha \in \mathbb{R}^L} J(\va)\\
    J : \va \mapsto \bar{m}\exp\left(\Phb\va\right) +  (\gamma \boldsymbol{1} -y)\Tr (\Phb \va) + 
    \gamma \norm{ \Phb\va - \mu}_{\Sigma^{-1}}^2 + \lambda \Psi(\va)~, \nonumber
  \end{gather}
} where $\boldsymbol{1}$ is the vector of ones, $\Sigma$ and $\mu$ the parameter of the log-normal prior and its
weighting parameter $\gamma$, $\Psi : \va \mapsto \sum_i \psi(\va[i])$ the sparsity-penalty and $\lambda$ the
regularization parameter. Notice that, we implicitly assume that the $\va[i], 0\le i \le n$ are independent and
identically distributed. Then the solution is given by $x = \exp(z) - 1 = \exp(\Phb\va) -1$.

%  As the change of variable
% is bijective, \eqref{eq:3} is somewhat equivalent to \eqref{eq:6}, but with the sparsity penalty over the underlying
% Gaussian field.

From $(\mathrm{P}_{\lambda,\gamma,\psi})$  we can characterize the solution,
{\small
  \begin{propo}
    {~} \\ \vspace{-0.5cm}
    \label{prop:objectives}
    \begin{enumerate}
      \setlength{\topsep}{0pt}
      \setlength{\parskip}{0pt}
      \setlength{\itemsep}{0pt}
      \setlength{\partopsep}{0pt}
    \item Existence:  $J \in \Gamma_0(\mathbb{R}^L)$, then $(\mathrm{P}_{\lambda,\gamma,\psi})$ has at least one solution.
    \item Uniqueness: $(\mathrm{P}_{\lambda,\gamma,\psi})$ has a unique solution if $\psi$ is strictly convex, i.e. if
      $\Phb$ is an orthobasis or if $\psi$ is strictly convex.
    \end{enumerate}
  \end{propo}
}

% Let $\Mc$ be the set of minimizers of problem \eqref{eq:3}. Suppose that $\Psi$ is coercive. Thus $J'$ is
% coercive. Therefore, the following holds:
% \begin{propo}
%  \begin{enumerate}
%     \setlength{\topsep}{0pt}
%     \setlength{\parskip}{0pt}
%     \setlength{\itemsep}{0pt}
%     \setlength{\partopsep}{0pt}
%   \item Existence: \eqref{eq:3} has at least one solution, i.e. $\Mc\ne\emptyset$.
%   \item Uniqueness: \eqref{eq:3} has a unique solution if $\Psi$ is strictly convex or if $\Phb$ is an orthobasis.
%   \end{enumerate}
% \end{propo}

\subsubsection{Solving the optimization problem}

We first define the notion of a proximity operator, which was introduced as a generalization of the
notion of a convex projection operator.
{\small
  \begin{definition}[{\cite{Moreau1962}}]
    \label{def:1}
    Let $\varphi \in \Gamma_{0}(\Rn)$. Then, for every $x\in\Rn$, the function $y \mapsto \varphi(y) + \norm{x-y}^{2}/2$
    achieves its infimum at a unique point denoted by $\prox_{\varphi}x$. The operator $\prox_{\varphi} : \Rn \to \Rn$
    thus defined is the \textit{proximity operator} of $\varphi$.
  \end{definition}
}
Then, the proximity operator of the indicator function of a convex set is merely its euclidean projector. With $\psi =
\abs{.}$, the proximity operator $\prox_{\lambda\Psi}$ is the popular soft-thresholding (denoted $\mathrm{ST}$) with threshold
$\lambda$.

%  We now turn
% to $\prox_{\eqi\psi}$ which is given by the following result:
% \begin{lemma}[{\cite{Combettes2011}}]
%   \label{th:3}
%   If $\psi = \abs{.}$, then the proximity operator, $\prox_{\eqi\Psi} : \va \mapsto \left( \sign(\va[i]) \max\{\abs{\va[i]} -
%     \lambda,0\} \right)_{0\le i \le L}$.
% \end{lemma}
% \begin{theorem}
%   \label{th:3}
%   Suppose that $\psi$ satisfies, (i) $\psi$ is convex even-symmetric , non-negative and non-decreasing on $[0,+\infty)$,
%   and $\psi(0)=0$. (ii) $\psi$ is twice differentiable on $\mathbb{R}\setminus \{0\}$. (iii) $\psi$ is continuous on
%   $\mathbb{R}$, it is not necessarily smooth at zero and admits a positive right derivative at zero $\psi^{'}_+(0) =
%   \lim_{h\to 0^+} \frac{\psi(h)}{h} > 0$. Then, the proximity operator $\prox_{\delta\psi}(\beta) = \hat{\va}(\beta)$ has
%   exactly one continuous solution decoupled in each coordinate $\beta_i$ :
%   \begin{equation}
%     \label{eq:10}
%     \hat{\va}_i(\beta_i) =
%     \begin{cases}
%       0 & \text{if } \abs{\beta_i} \le \delta\psi^{'}_+(0)\\
%       \beta_i-\delta\psi^{'}(\hat{\va}_i) & \text{if } \abs{\beta_i} > \delta\psi^{'}_+(0)
%     \end{cases}
%   \end{equation}
% \end{theorem}
% A proof of this theorem can be found in \cite{Fadili2006}. Among the most popular penalty functions $\psi$ satisfying
% the above requirements, we have $\psi(\va_i) = \abs{\va_i}$, in which case the associated proximity operator is
% soft-thresholding (denoted $\mathrm{ST}$). 

If the dictionary $\Phb$ is a tight frame, then the proximal operator of its composition with a convex function $f$ is,
{\small
\begin{lemma}[{\cite{Combettes2008}}]
  \label{th:compo}
  If $\Phb$ is a tight frame,i.e. $\Phb\Phb\Tr = \nu\Id$, then $f \circ \Phb \in \Gamma_0(\Rn)$ and
  \begin{equation}
    \label{eq:12}
    \prox_{f\circ\Phb} = \Id + \nu^{-1} \Phb\Tr \circ (\prox_f - \Id) \circ \Phb
  \end{equation}
\end{lemma}
}

We also need the proximity operator from the terms of both data fidelity and log-normal prior.
{\small
  \begin{lemma}
  \label{th:expp}
  The proximity operator associated to $F : x \mapsto \bar{m}\exp(x) + (\eqi\boldsymbol{1} - y)^Tx + \gamma \norm{x -
    \mu}_{\Sigma^{-1}}^2$ is,
   \begin{gather}
      \prox_{\beta F} x  = \Kmb^{-1} \prox_{\beta \bar{m}\exp} \big( \Kmb^{-1}\big(x + \beta \left(y - \eqi\boldsymbol{1} + \eqi \beta \Sigma^{-1}\mu \right) \big) \big)~, \nonumber \\
      \label{eq:5}  \text{with }\prox_{\beta\bar{m} \exp}  x = \log\left( \Wm(\beta\bar{m} \exp(x))/(\beta\bar{m}) \right)~,
    \end{gather}
    where $\Wm$ is the Lambert $\Wm$ function~\cite{Corless1996} and $\Kmb = \Id + \eqi \beta \Sigma^{-1}$.
\end{lemma}
}
% This lemma comes from both the properties of proximity operator~\cite[Lemma 2.6]{Combettes2005} and the properties of
% the Lambert $\Wm$ function~\cite{Corless1996}.

Then, we propose to use the generalization of the Douglas-Rachford algorithm presented in \cite{Combettes2008} in order
to solve \eqref{eq:3}. The solution is computed using the iterative scheme presented by Algorithm~\ref{algo:denoise}.

\begin{algorithm}[h]
  \small
  \noindent{\bf{Task:}} Estimate the density field. \\
  \noindent{\bf{Parameters:}} The observed image counts $y$, the mean number of count $\bar{m}$, the dictionary $\Phb$,
  the number of iterations $N_{\mathrm{est}}$, the proximal step $\mu$, the log-normal prior parameter $\Sigma$ and
  $\mu$, and the regularizations parameters $\lambda$ and $\eqi$. \\
  \noindent{\bf{Initialization:}}\\
  $\forall i \in \{0,1\},\quad p_{(0,i)} = \Phb\Tr y$. \\
  $\va_0 = \Phb\Tr y$. \\
  \noindent{\bf{Main iteration:}} \\
  \noindent{\bf{For}} $t=0$ {\bf{to}} $N_{\mathrm{est}}-1$,
  \begin{itemize}
    \setlength{\topsep}{0pt}
    \setlength{\parskip}{0pt}
    \setlength{\itemsep}{0pt}
    \setlength{\partopsep}{0pt}
  \item \underline{Data fidelity with log-normal prior} (Lemma~\ref{th:compo} and \ref{th:expp}): 
    $\xi_{(t,0)} = p_{(t,0)} + c^{-1} \Phb\Tr \circ ( \prox_{\mu F/2} - \Id ) \circ \Phb (p_{(t,0)})$.
  \item \underline{Sparsity-penalty}: $\xi_{(t,1)} =
    \prox_{\mu \lambda\Psi/2} p_{(t,1)} = \mathrm{ST}_{\mu\lambda/2}(p_{(t,1)})$.
 \item Average the proximity operators: $\xi_{t} = (\xi_{(t,0)} + \xi_{(t,1)})/2$. 
  \item Choose $\theta_t\in]0,2[$.
  \item Update the components: $\forall i \in \{0,1\},\quad p_{(t+1,i)} = p_{(t,i)} + \theta_t (2\xi_{t} - \va_{t} - \xi_{(t,i)})$.
  \item Update the coefficients estimate: $\va_{t+1} = \va_t + \theta_t(\xi_t - \va_t)$
  \end{itemize}
  \noindent{\bf{End main iteration}} \\
  \noindent{\bf{Output:}} Denoised field $\delta^{\star}=\exp(\Phb\va_{N_{\mathrm{est}}}) - 1$.
  \caption{Density field estimation, solve $(\mathrm{P}_{\lambda,\gamma,\psi})$.}
  \label{algo:denoise}
\end{algorithm}

\section{Results on the 2MASS survey}
\label{sec:results}

As an experiment, we apply our method on the 2MASS galaxy survey~\cite{Jarrett:2004a}. As we are working on the sphere,
the dictionary $\Phb$ contains the spherical harmonics transform, which is an orthobasis.  The manually tuned parameters
were $N_{\mathrm{tex}}=15$, $N_{\mathrm{est}}=40$ ,$N_{\mathrm{iter}}=6$, $N_{\mathrm{MI}}=10$, $\eqi = 10^{-4}$ and
$\lambda = 10^{-3}$, that seems sufficient for recovering most of the large scales. This method was compared with the
inpainting method proposed in \cite{inpainting:abrial06} (denoted M2) which fills in the missing data area using both
sparsity and a quadratic data fidelity.

\begin{figure}[htb]
  \begin{center}
    \begin{tabular}{c}
      \includegraphics[width=0.6\linewidth]{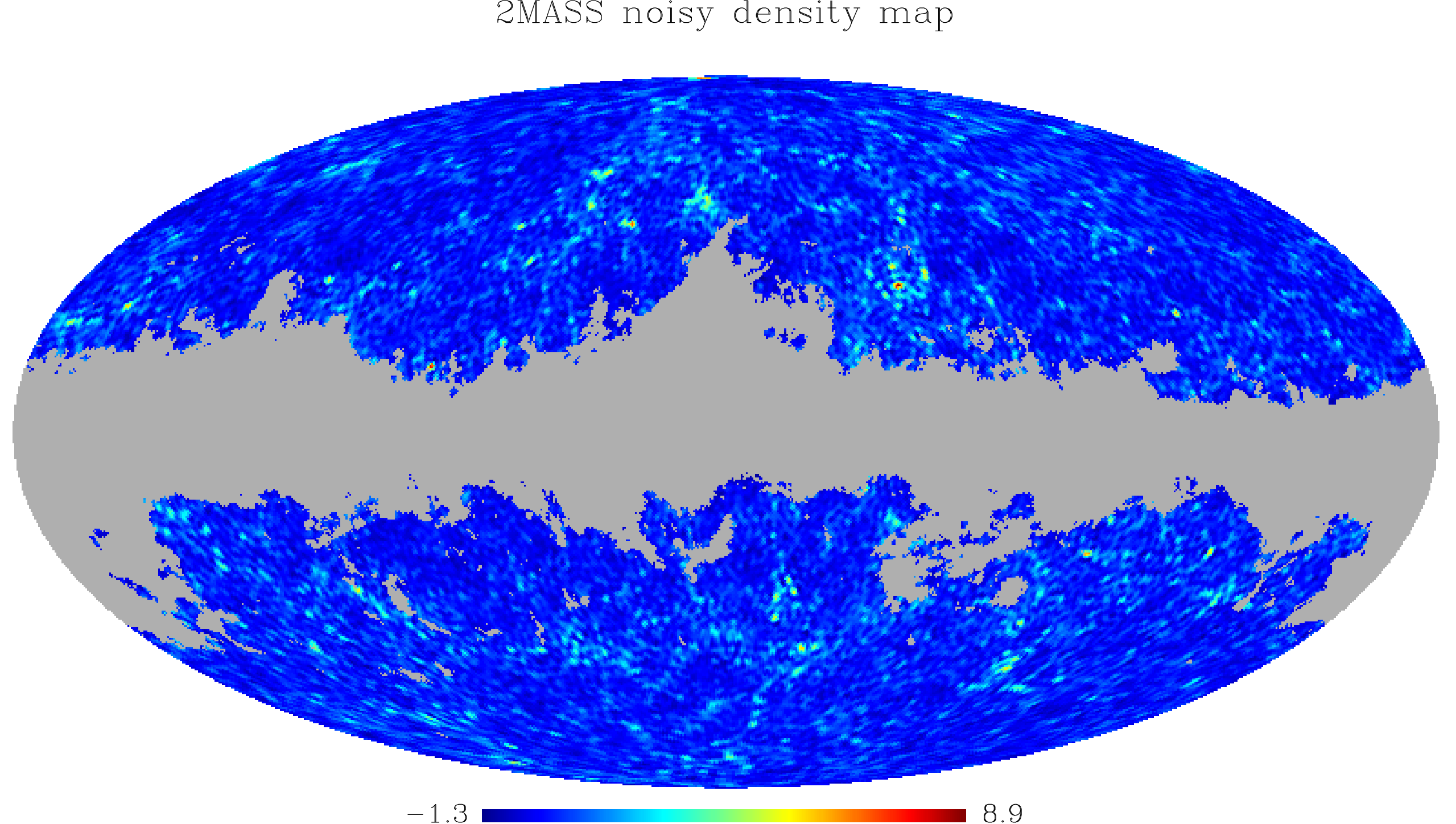} \\
      \includegraphics[width=0.6\linewidth]{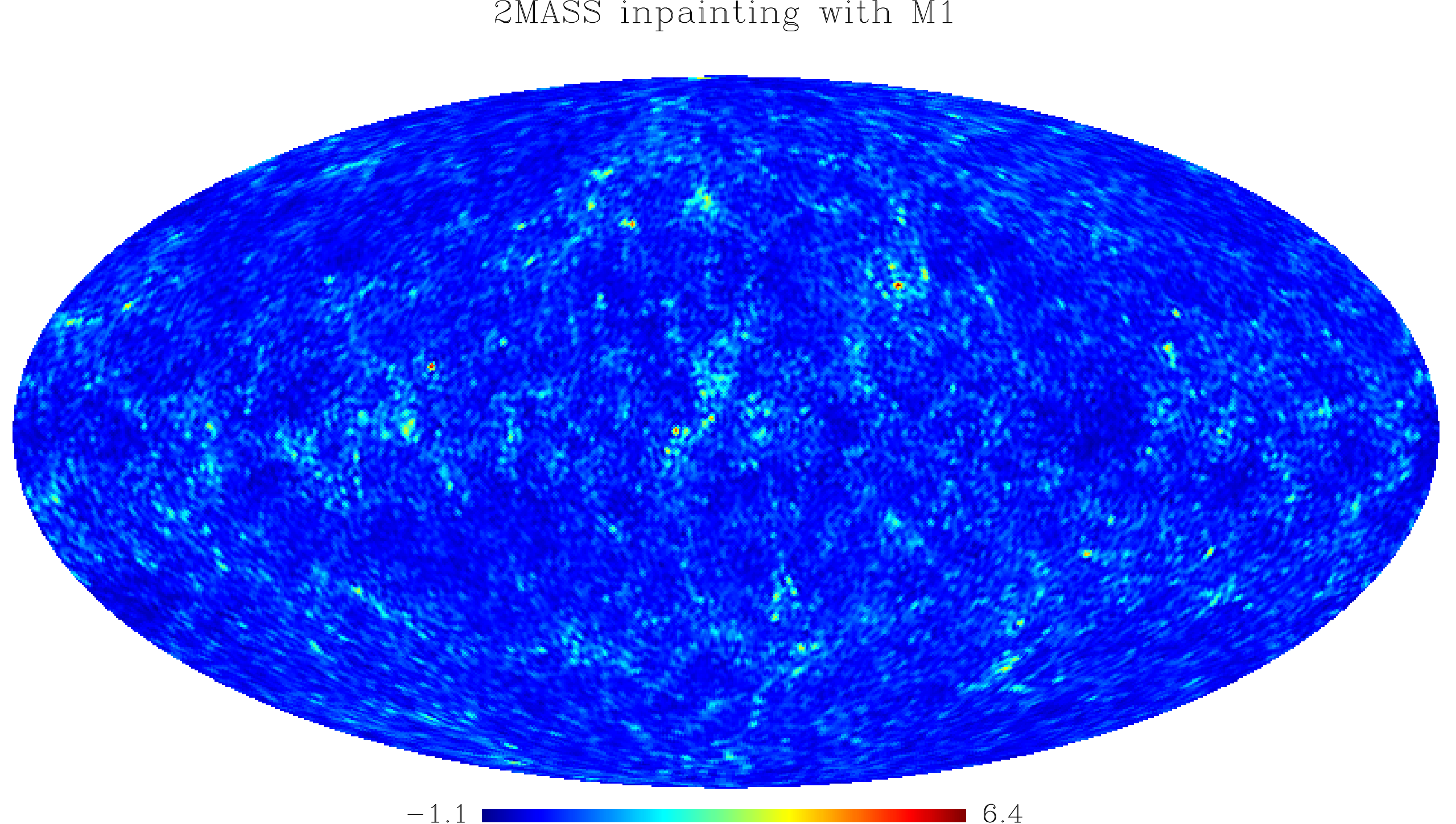} \\
      \includegraphics[width=0.6\linewidth]{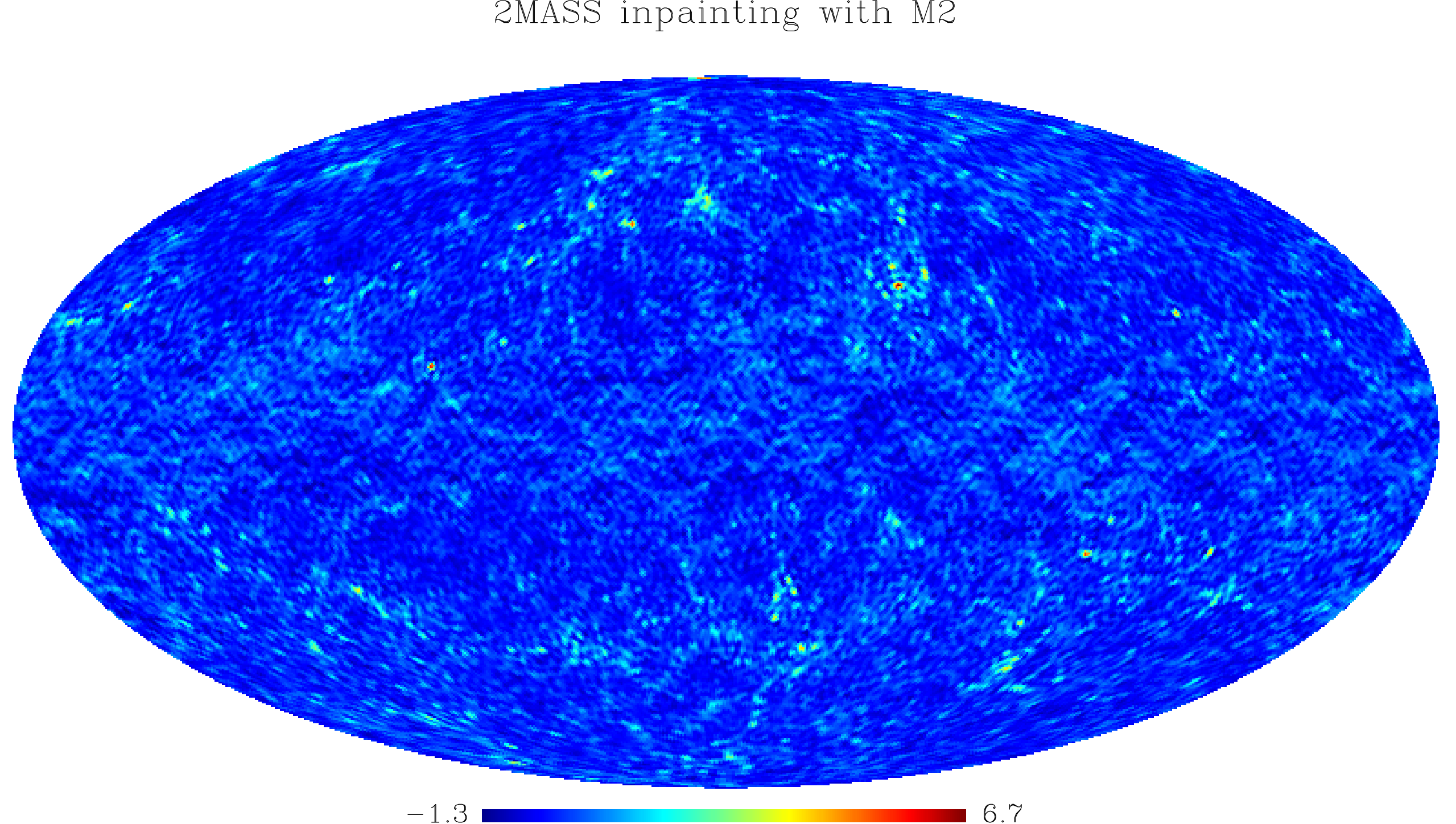} \\
    \end{tabular}
  \end{center}
  \caption{Results on the inpainting methods on the 2MASS count map. Top: The 2MASS noisy density map with the missing
    data in gray. Middle and bottom: The inpainted density map using the method M1 (Middle) and M2 (bottom).}
  \label{fig:2mass}
\end{figure}

The results are pictured by Fig.~\ref{fig:2mass}. In order to compare efficiently the results, we remove all the
spherical harmonic modes beyond $200$ (i.e. $\ell \le 200$) of the maps. Method M1 gives a better estimation of the
inpainted areas, as realistic structures has been created inside these areas and the transition between missing and
observed pixels are invisible. While with method M2, transitions can be clearly seen and no structure is infer inside
the large missing area in the center.

\begin{figure}
  \begin{center}
    \includegraphics[width=0.4\linewidth]{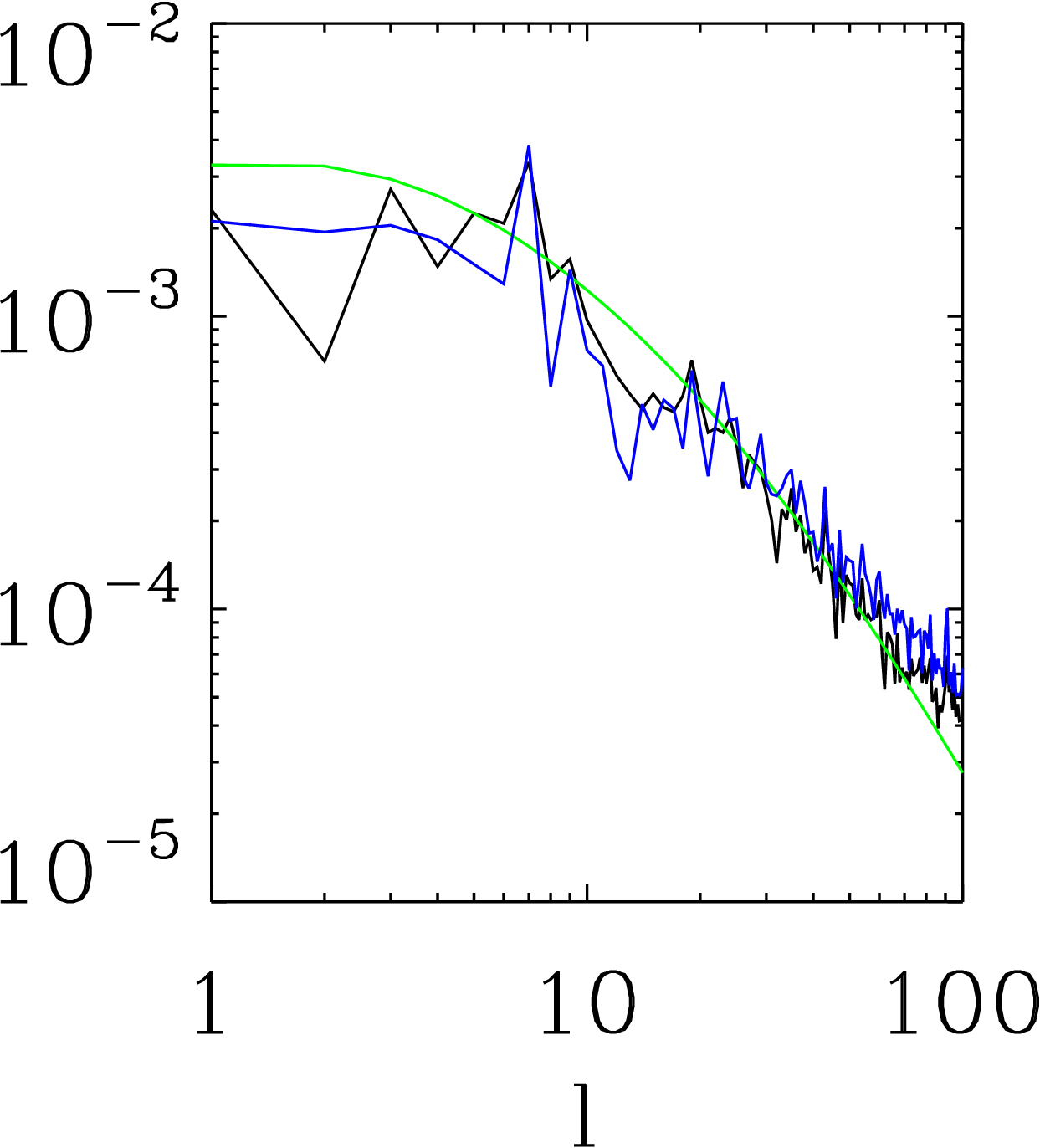}
  \end{center}
  \vspace*{-0.8cm}
  \caption{The theoretical (green), M1 (black) and M2 (blue) power spectra for the first 100 harmonics modes.}
  \label{fig:powsp}
  \vspace*{-0.4cm}
\end{figure}

For the denoised zones, both methods preserve the structures and the amplitude. In order to compare the behavior of the
two methods, we also compare in Fig.~\ref{fig:powsp} the second-order statistics of the inpainted maps to the
theory~\cite{Afshordi:2003xu}.  More precisely, we focus on the first modes of the harmonic power spectrum of the
density field.  While at the beginning methods M1 and M2 provide similar results, they differ on higher modes where the
Poisson noise becomes more salient at the profit of the M1 method.

%   The power spectra of the
% inpainted maps are showed in Fig~\ref{fig:powsp} and compared to the theory~\cite{Afshordi:2003xu}. Notice while M1 and
% M2 are very similar for the largest scales, they begin to differ on higher scales where the Poisson noise becomes more
% salient.

\section{Conclusion}
\label{sec:conclusion}
\vspace*{-0.2cm}

An inpainting method is proposed using a data augmentation procedure, where the observation is first completed with
realistic data. For galaxy density field, we propose to use two priors, first we assume that the density field follows a
log-normal distribution and secondly, the underlying Gaussian field is assumed to be sparse inside a wisely chosen
dictionary. The resulting algorithm is able to preserve the second order statistical properties which is an important
feature for astrophysics application.


{ \footnotesize
  \bibliographystyle{IEEEbib}
  \bibliography{references}
}
% \bibliography{strings,refs}

\end{document}